\documentclass[
,twocolumn%
,secnumarabic%
,showpacs,showkeys
,tightenlines,amssymb, amsmath,nobibnotes, nofootinbib,aps, prl]{revtex4}
\usepackage[dvips]{graphicx}
\usepackage{tabularx}
\usepackage{bm}%
\expandafter\ifx\csname package@font\endcsname\relax\else
 \expandafter\expandafter
 \expandafter\usepackage
 \expandafter\expandafter
 \expandafter{\csname package@font\endcsname}%
\fi

\begin{document}
\title{The Zero Temperature Phase Diagram of the Kitaev Model}

\author{Charles Nash}
\email[]{cnash@thphys.nuim.ie}
\affiliation{Department of Mathematical Physics, NUIM, Maynooth, Kildare, Ireland.}
\author{Denjoe O'Connor}
\email[]{denjoe@stp.dias.ie}
\affiliation{School of Theoretical Physics,
DIAS, 10 Burlington Road, Dublin 4, Ireland. }

\begin{abstract}
We show that the zero temperature phase diagram of the vortex free
sector of the Kitaev model is in one to one correspondence with that
of the classical dimer model on the same lattice. We find
that the model generically has three distinct phases.  On a honeycomb
lattice with a $3\times3$ fundamental domain all three phases are
accessible. As the couplings are varied there are two distinct
transitions. The new transition is one to a gapped phase that opens up
in the interior of the $B$ phase.

\end{abstract}
\pacs{75.10.Jm, 73.43-f, 75.50.Mm, 05.50.+q,05.30.Pr}
\keywords{Chiral Spin Liquid, Kitaev Model, Dimer Models, Vortices}

\maketitle

{\it Introduction.}---Kitaev's observation \cite{Kitaev:2005} that a
spin $1/2$ system on a honeycomb lattice has a gapless phase with
vortex excitations that obey non-Abelian statistics has stirred a lot
of
interest \cite{Pachos:2006,Feng_Zhang_Xiang:2007,Maynooth_colleagues,Yao_Kivelson,ChenNussinov:2008,SchmidtDusuelVidal:2008,DusuelSchmidtVidalZaffino:2008,LeeZhangXiang:2007}. Yao
and Kivelson \cite{Yao_Kivelson} extended Kitaev's considerations to
the Fisher (triangle-honeycomb) lattice and found that its
zero-temperature ground state is a chiral spin liquid.  We extend
these considerations to the more general setting.

Kitaev models divide into two classes: those defined on bipartite and
non-bipartite lattices. A lattice is bipartite if its sites can be
coloured black and white so that adjacent sites are always of the
opposite colour. A Kitaev model is then bipartite if its Hamiltonian has 
no interaction between bonds of the same colour. We establish that 
the vortex free, zero temperature, phases of Kitaev models are in 
one to one correspondence with the phases of classical dimer models.
\begin{itemize}
\item
Bipartite models have generically three phases, the third being 
a previously unnoticed gapped phase which we refer to as the $C$ 
phase. This arises {\it only} when a sufficiently large fundamental domain 
is considered, e.g. a $3\times3$ hexagonal domain.
\end{itemize}
As the couplings are varied the 
$C$ phase opens up in the interior of the $B$ phase 
and corresponds to giving a Dirac mass to the pair of degenerate massless 
Majorana Fermions of the $B$ phase. 
Close to the transition when the mass gap is small the model 
is well approximated by a continuum massive Dirac field theory.
\begin{itemize}
\item
Non-bipartite couplings lift the degeneracy of the Majorana modes
and the mass gap of one of the Majorana fields can go to zero to give
a chiral spin liquid as discussed in \cite{Yao_Kivelson}.
\end{itemize}

After summarizing the Kitaev model and its formulation in terms of
Majorana operators we establish that the quadratic form associated
with the Hamiltonian in the vortex free phase is precisely the
Kasteleyn matrix of a classical dimer model. This allows us to
identify the phase diagrams of the two models and establish the
existence of a new gapped $C$ phase in the bipartite Kitaev model and
predict the mass gap. It also affords us an understanding of
non-bipartite models such as the extended Kitaev model
of \cite{Yao_Kivelson} where the couplings can
be choosen to correspond to those of a classical Ising model in its
dimer realization \cite{Fisher:1966}. At the Ising critical point the
model has a low-lying massless spectrum that corresponds to a
``relativistic'' chiral Fermion \cite{Haldane:1988} which becomes
massive as the couplings are varied.


{\it The model.}---The Kitaev spin model on the hexagonal lattice is
given by the Hamiltonian
\begin{equation}
{\cal H}=\sum_{x-{\rm link}}J_x\sigma^x_i\sigma^x_j
+\sum_{y-{\rm link}}J_y\sigma^y_i\sigma^y_j
+\sum_{z-{\rm link}}J_z\sigma^z_i\sigma^z_j
\label{Kitaev_Hamiltonian}
\end{equation}
and partition function
\begin{equation}
Z=Tr({\rm e}^{-\beta {\cal H}}) . 
\label{quantum_part_fn}
\end{equation}
For simplicity we only consider the case of open boundary conditions
as our focus is on the phase diagram of such models.

With open boundary conditions the model (\ref{Kitaev_Hamiltonian}) 
can be rewritten, after a Jordan-Wigner transformation 
\cite{Kitaev:2005,Feng_Zhang_Xiang:2007,Yao_Kivelson}, 
in terms of Majorana operators
$c_i$ with the partition function (\ref{quantum_part_fn}) given by
\begin{equation}
Z=\sum_{u}Tr({\rm e}^{-\beta H(u)})
\end{equation}
and the  
Hamiltonian 
\begin{equation}
H(u)=\frac{i}{2}
\sum_{ij}
c_iK_{ij}(u)c_j
\end{equation}
where $\{c_i,c_j\}=2\delta_{ij}$.  The matrix element $K_{ij}$ between 
site $i$ and $j$ is given by one of the couplings $J_x$, $J_y$ or $J_z$, 
up to an overall sign.   
This rewriting is valid for quite general couplings which 
can vary throughout the lattice. 
Yao and Kivelson \cite{Yao_Kivelson} extended Kitaev's 
construction to the Fisher lattice (or triangle-honeycomb 
lattice).

The most general Kitaev model then assigns positive 
couplings to all links and a $Z_2$  gauge field $u_{ij}$
to each $z$-link between sites $i$ and $j$. 

Unfortunately, for general temperatures 
the Kitaev model has not yet yielded an exact solution. However,
the zero temperature ground state is known, thanks to Lieb's theorem
\cite{Lieb:1994}, to be the vortex free state.

{\it Phase diagram equivalence.}---In the vortex free case
the $u_{ij}$ become $u^{\rm std}_{ij}$, and $K_{ij}$ is the Kasteleyn
matrix of a classical dimer model. In this context $u^{\rm std}_{ij}$
is a Kasteleyn orientation given by the ``clockwise odd
rule'' \cite{Fisher:1966,Kasteleyn:1967} around a plaquette.  In the
thermodynamic limit the eigenvalues of $K$ can be used to map out the
phase diagram of the model. The phases of the zero temperature Kitaev
model are therefore in one to one correspondence with those of
classical dimer models.

Dimer models divide into two classes: those defined on bipartite and
non-bipartite lattices. The hexagonal lattice is bipartite while the 
Fisher lattice is non-bipartite.
The physics of the bipartite and non-bipartite lattices are quite
distinct, so we will discuss them separately.

We begin with the bipartite case for which one can write $K$ in the form
\begin{equation}
K=\left(\begin{array}{cc}
                  0&A \cr 
                   - A^T&0\cr
            \end{array}\right)
\label{Kasteleyn}
\end{equation}
where the matrix $A$ has real entries and maps between the 
black and white sites.

Fourier transforming $K$ one obtains a $2d\times 2d$ matrix where $d$
is the number of sites of the same colour in a fundamental domain and whose 
determinant $D({\bf k})=\vert P\vert^2$ where 
$P={\rm Det}[A({\bf k})]$ with ${\bf k}=(\theta,\phi)$. 

$P$, the determinant of $A$, is then a (Laurent) polynomial 
$P({\rm e}^{i\theta},{\rm e}^{i\phi})$ in two unimodular complex variables.
One can move off the unit circle by using some of the couplings to
define general complex variables $z$ and $w$.  We will refer the resulting 
polynomial as the spectral polynomial of the model. It can generically 
be written in the form
\begin{equation}
P(z,w)=\sum_{a,b}{(-1)}^{a+b+ab}p_{ab} z^a w^b
\end{equation}
where $a$ and $b$ are integers and the coefficients 
$p_{ab}$ are positive real numbers determined by the couplings of the model.

The zero locus,
\begin{equation}
P(z,w)=0\ ,
\label{spectral-curve}
\end{equation}
allows one to immediately identify the gapless phase $D({\bf k})=0$.
This zero locus is called the spectral curve.

Kenyon, Okounkov, and Sheffield \cite{Kenyon:2003uj} and Kenyon and
Okounkov \cite{Kenyon:2003ui} proved that the spectral curve
(\ref{spectral-curve}), of a $d\times d$ fundamental domain with
generic couplings is a Harnack curve and furthermore that every
Harnack curve arises as the spectral curve of some bipartite dimer
model.  Such curves arose in the early twentieth century mathematics
literature.  They have very special properties and can be
characterized by what is referred to as the {\it amoeba} of the curve.

Given the correspondence of the quadratic form of the vortex free sector 
of the Kitaev model to the Kasteleyn matrix of a classical dimer model, 
as pointed out above, eqn. (\ref{spectral-curve}), determines the phase 
diagram of the zero vortex sector of the Kitaev model. 
This phase diagram can conveniently be
plotted in terms of the amoeba of eqn. (\ref{spectral-curve}) as 
we now describe.

Put simply, with
$z={\rm e}^{x+i\theta}$ and $w={\rm e}^{y+i\phi}$,
eqn. (\ref{spectral-curve}) can be solved for the angles
$\theta$ and $\phi$ (see \cite{Nash_OConnor_jphysa:2009}). There 
are only two solutions
$(\Theta(x,y),\Phi(x,y))$ and $(-\Theta(x,y),-\Phi(x,y))$ 
corresponding to $(z,w)$ and $(\bar{z},\bar{w})$. 
At the  boundary of the zero locus these angles become either $0$ 
or $\pi$. The interior of the domain in the $x$, $y$ plane bounded by
these curves is the {\it amoeba}.

In brief the amoeba can be plotted by noting that on it 
\begin{equation}
\prod_{p,q=0}^{1}P({\rm e}^{x+i p\pi},{\rm e}^{y+iq\pi})\leq0\ .
\end{equation}
The amoeba has bounding curves called {\it ovals}. Ovals
never intersect and can be both compact and non-compact. 
The genus, $g$, of the spectral 
curve is equal to the number of compact ovals. 

\begin{figure}[t]
 \begin{center}
 \includegraphics[height=30mm]{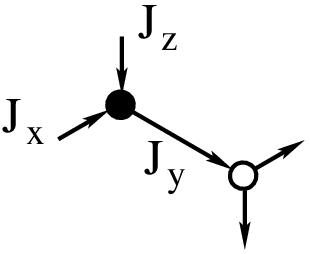}
\hfill\includegraphics[height=30mm]{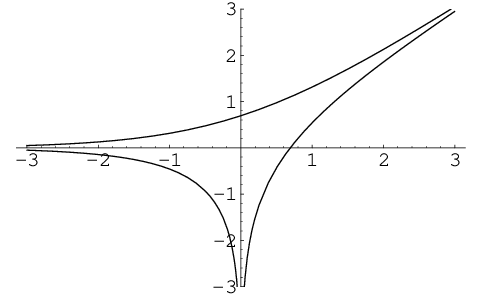}
 \caption{\footnotesize\label{fig:hex_simple} 
On the left is the basic tile for the 
hexagonal lattice, showing the activities and our choice of Kasteleyn orientation, with the
corresponding amoeba on the right. The regions exterior to the curves shown 
corresponds to the Kitaev $A$ phase while the interior or amoeba 
gives the $B$ phase.}
\end{center}
 \end{figure}

The {\it phase diagram} of both classical bipartite dimer models and
the zero temperature, zero vortex sector of Kitaev models then
consists of the amoeba and its complement in the plane. The phase
transitions of these models occur on the boundary of the amoeba,
i.e. on the ovals.

For the $1\times 1$ fundamental domain with positive 
couplings $J_x$, $J_y$ and $J_z$ 
the spectral curve is given by 
\begin{equation}
P(z,w)= 1-1/z-w=0
\end{equation}
where $z=(J_y/J_x) {\rm e}^{i\theta}$ and $w=(J_z/J_y) {\rm e}^{i\phi}$.
We show the fundamental tile and its corresponding amoeba in 
Fig. \ref{fig:hex_simple}, see \cite{Nash_OConnor_jphysa:2009} for more 
details on the amoeba in this case.

The complement of the amoeba consists of both compact and non-compact 
regions. In the terminology of dimer models, 
as models of melting crystals, the
non-compact regions exterior to the bounding ovals constitute the
frozen regions. The amoeba itself is referred to as the liquid phase,
and the interior of the compact ovals as the gaseous phase.

The amoeba corresponds to the critical surface and
gives the gapless $B$ phase of the Kitaev model. The frozen dimer
regions go over to the Kitaev $A$ phase.  There is however a 
{\it  third $C$ phase} corresponding to the gaseous phase of the dimer model.

{\it A gapped phase.}---This $C$ phase seems not to have been observed
in the literature. It is a novel phase in that it has a gap which
corresponds to a Dirac mass for the pair of Majorana Fermions that
describe the low energy excitations of the model.

\begin{figure}[t]
 \begin{center}
 \includegraphics[height=30mm]{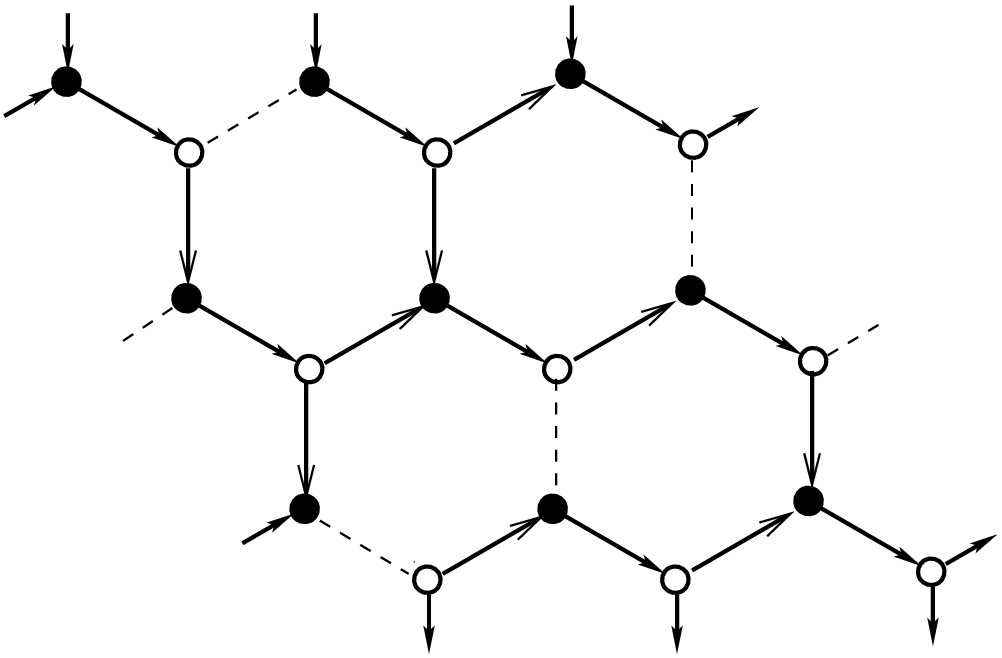}
 \caption{\footnotesize\label{fig:square_equiv-hex33} 
The hexagonal tiling which reduces to the square lattice on bond 
contraction. Dashed lines correspond to bonds that are set to zero
while all other bonds are allowed to have generic values.}
\end{center}
 \end{figure}

\begin{figure}[t]
 \begin{center}
 \includegraphics[height=30mm]{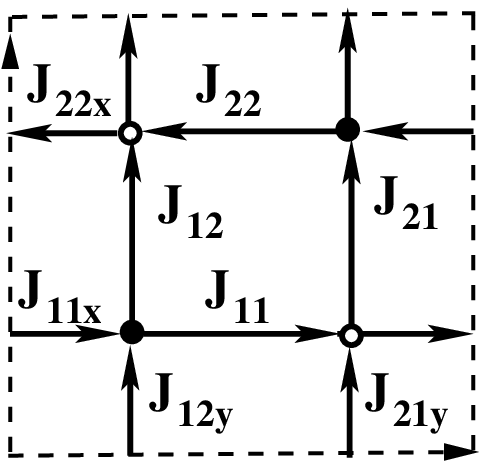}
\hfill \includegraphics[height=30mm]{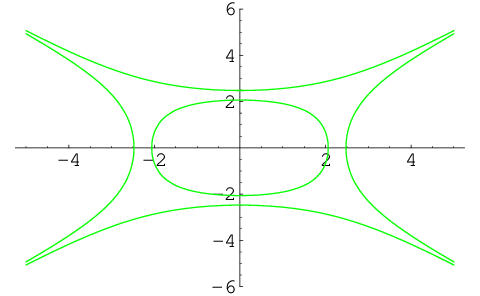}
 \caption{\footnotesize\label{fig:square_general} 
The most general square lattice tiling
with two vertices of each colour per tile and its amoeba 
with $t_z=t_w=0$ and $A=B=1$ and $D=10$ in eqn. (\ref{Psquare_zw}).}
\end{center}
 \end{figure}

The simplest example of a model that exhibits a $C$ phase is the
square lattice as shown in Fig. \ref{fig:square_general}. This can be
obtained, by bond contractions \cite{Fisher:1966,Kenyon:2003ui}, from
a hexagonal lattice, with a $3\times3$ fundamental cell, as shown in
Fig. \ref{fig:square_equiv-hex33}. The Hilbert space of
(\ref{Kitaev_Hamiltonian}) then grows as $2^{18n}$ with $n$ the
number of copies of the fundamental cell and makes direct or Monte Carlo
analysis of the phase diagram difficult.

The Kasteleyn matrix takes the form (\ref{Kasteleyn}) and for the square lattice
with four site fundamental tile, which is the
simplest model with a compact oval, we have
\begin{equation}
A=\left(\begin{array}{cc}
                  J_{11}-J_{11x}{\rm e}^{-i\theta}& J_{12}-J_{12y}{\rm e}^{-i\phi}\cr 
                   -J_{21}+J_{21y} {\rm e}^{i\phi}&J_{22}-J_{22x} {\rm e}^{i\theta}\cr
            \end{array}\right)
\label{GeneralSquare}
\end{equation}
The determinant of $A$ gives $P^{Square}$
which can be parameterized in the form
\begin{equation}
P^{{\rm Square}}=D-A(z+\frac{1}{z})
-B(w+\frac{1}{w})
\label{Psquare_zw}
\end{equation}
where $D=2A\cosh(t_x)+2B\cosh(t_y)$ and 
$z={\rm e}^{x+i\theta}$ and $w={\rm e}^{y+i\phi}$ so that
$z$ and $w$ can now take any complex values.
The key point here is that this particular model has three phases:
Those described in Kitaev's paper \cite{Kitaev:2005} as the $A$ and $B$ phases 
together with a new $C$ phase involving a mass gap. This phase arises when 
$t_x$ or $t_y$ is greater than zero. The phase is bounded by the curve
$P^{{\rm Square}}({\rm e}^{x},{\rm e}^{y}) =0 $ which is the compact oval.

Setting $J_{11x}={\rm e}^{-t}, J_{22}={\rm e}^{t}$ and all others to
unity puts the model in its $C$ phase and corresponds to the origin of
Fig. \ref{fig:square_general} where $t=\ln(4+\sqrt{15})$.  The masses
of the Majorana modes are $m_1=1-{\rm e}^{-t}$ and $m_2={\rm
e}^{t}-1$. For small $t$ the mass gap is given by $t$ and vanishes
when $t=0$ where the $C$ phase disappears. More generally the gap
can be obtained from the lowest eigenvalue of the Kasteleyn matrix, or
estimated from the spectral polynomial whose modulus squared is the
determinant of $K$.

The general bipartite lattice---all of which can be
obtained by special reductions, involving bond contractions
\cite{Fisher:1966,Kenyon:2003ui},
of the hexagonal lattice with $d\times d$ fundamental domain
containing $d^2$ vertices of each colour---is as follows: The phase
diagram of the zero temperature, zero vortex Kitaev model is specified
by the amoeba of the spectral polynomial which is given as the
determinant of $A$ as in eqn. (\ref{Psquare_zw}).  The Kasteleyn matrix,
$K$, then describes a set of Majorana Fermions. On the amoeba the
system is gapless with two of these Fermions being massless. These in
turn become massive as one moves inside a compact oval and the system
becomes gapped.

There will in fact be $g=(d-1)(d-2)/2$ compact ovals.  The number,
$g$, of compact ovals is a topological invariant and is the genus of
the spectral curve. As the couplings are varied some of the ovals may
contract to points.  When bond couplings are set to zero, bounding
phases will disappear.  One can remove the corresponding bond from the
tiling.

When the system is wrapped on a torus there is an additional term 
in the Hamiltonian \cite{Yao_Kivelson} and a projection operator in 
the expression for the partition function (\ref{quantum_part_fn}). 
However, one can make the following observation: The spectrum of the Kasteleyn matrix, $K$, is 
discrete and on the amoeba the partition function depends on the shape 
parameter, $\tau$, (often referred to as the modular parameter) of the torus. 
As one approaches the non-compact boundary of the amoeba the torus 
becomes degenerate and there is a topological transition where the volume of
the torus goes to zero \cite{Nash_OConnor_jphysa:2009}.
This can best be seen in the context of classical dimers
where the finite size corrections correspond to the conformal field theory
of a massless Dirac Fermion.

{\it Non-bipartite models.}---The classical dimer model is also useful
in gaining an understanding of the Kitaev model on a non-bipartite
lattice. Again the minimum energy configuration will correspond to
$u^{{\rm std}}$ being a Kasteleyn orientation and $K(u^{{\rm std}})$ being a
Kasteleyn matrix and its critical surface
will describe the gapless phase.

\begin{figure}[t]
\begin{center} 
\includegraphics[height=35mm]{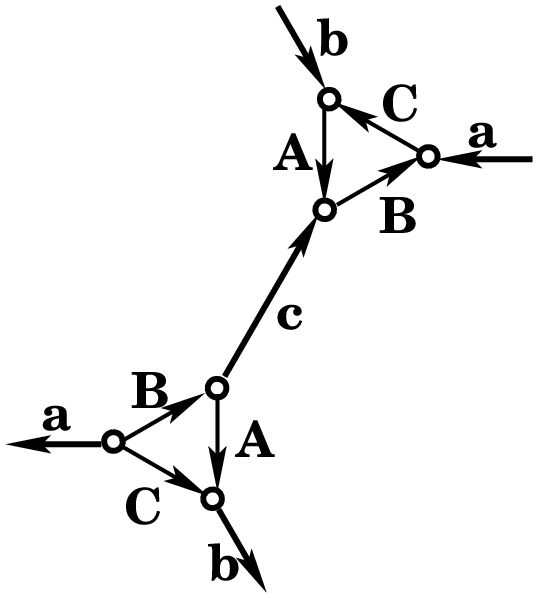} 
\caption{\footnotesize\label{fig:Fisherlattice}
 The Fisher Lattice.}
\end{center}
 \end{figure}

A classic example is the dimer model on the Fisher lattice, 
Fig. \ref{fig:Fisherlattice}, whose Kasteleyn matrix is given by
\begin{equation}
K=\left(\begin{array}{cccccc}
                  0&0&0&c&-B& A\cr 
                  0&0&C& -B&-a {\rm e}^{i\theta}&0\cr
                  0&-C&0 &A&0&-b {\rm e}^{i\phi}\cr
                  -c&B&-A&0&0&0\cr 
                   B&a {\rm e}^{-i\theta}&0&0&0&C\cr
                   -A&0&b {\rm e}^{-i\phi}&0&-C&0\cr
            \end{array}\right) . 
\label{KasteleynFisher}
\end{equation}
The Pfaffian of $K$ gives the partition function of the two
dimensional Ising model \cite{Fisher:1966}.  Setting $A$, $B$ or $C$
to zero renders the lattice bipartite and equivalent, by a lattice
reduction \cite{Fisher:1966,Kenyon:2003ui}, to the hexagonal
lattice. When $C=0$
\begin{equation}
P^{{\rm Fisher}}=-a A^2 {\rm e}^{i\theta}- b B^2 {\rm e}^{i\phi} 
-abc {\rm e}^{i(\theta+\phi)} .
\end{equation}
This crossover from bipartite to non-bipartite Ising like behaviour
was studied in \cite{Bhattacharjee:1984}.

One can apply this result to the Kitaev model
on the Fisher lattice, studied in \cite{Yao_Kivelson}. 
Off the Ising critical surface the Kitaev model is gapped 
and is gapless at the critical point. 
The mass of the lightest Majorana Fermion gives the gap.

Setting $A=B=C=1$ and $a=1/v_1$, $b=1/v_2$ and $c=1/v_3$, with 
$v_i=\tanh(k_i)$, the classical dimer model gives the partition function 
\begin{equation}
Z_{{\rm Ising}}^{{\rm Hex}}=C_{{\rm norm}}Z_{{\rm dimer}}^{{\rm Fisher}}(v_1^{-1},v_2^{-1},v_3^{-1},1,1,1)
\end{equation}
for the two dimensional Ising model 
on the hexagonal lattice. The normalization constant $C_{{\rm norm}}$ is given by
\begin{equation}
C_{{\rm norm}}=2^{2NM}{[\sin(k_1)\sin(k_2)\sin(k_3)]}^{NM}
\end{equation}
and $Z_{{\rm dimer}}^{{\rm Fisher}}$ is the dimer partition function on the Fisher lattice. 

The hexagonal lattice Ising model (and hence the Kitaev model)
is critical for $v_1v_2+v_2v_3+v_3v_1=1$; e.g. for  
$a=b=c=\sqrt{3}$ with $A=B=C=1$, the model is
gapless with two of the six eigenvalues zero at
$\theta=\phi=\pi$, in agreement with \cite{Yao_Kivelson}.

{\it Summary.}---In this note we established an equivalence of the
phase diagrams of the vortex free phases of Kitaev models with those
of classical dimer models. This allowed us to draw the following
conclusions: For bipartite models the $B$ phase, described by a
massless Dirac Fermion, corresponds to the amoeba of a classical dimer
model.  Changing the parameters of a sufficiently complicated model
one can cross a phase boundary to a new $C$ phase where this Fermion
acquires a mass. Adding non-bipartite couplings lifts the mass gap
degeneracy of the two Majorana Fermions. The zero-temperature ground
state is then a chiral spin liquid \cite{Yao_Kivelson}.  When the mass
gap corresponding to the lightest Majorana Fermion is very small the
zero vortex sector is well described by a relativistic chiral Fermion.

This work was partly supported by EU-NCG Marie Curie Network
No. MTRN-CT-2006-031962.

\end{document}